\documentclass[twocolumn]{aastex631}
\usepackage{soul}
\usepackage{tikz}
\usepackage{graphicx}
\usepackage{soul,verbatim}
\usepackage{hyperref} 

\newcommand{\dproj}{\ensuremath{d_{\rm proj}}}
\newcommand{\dv}{\ensuremath{\Delta v_{\rm los}}}
\newcommand{\Ha}{{H$\alpha$}}
\newcommand{\vmax}{\ensuremath{v_{\rm max}}}

\begin{document}
\title{Distribution of \Ha\ Emitters in Merging Galaxy Clusters}

\author[0000-0002-0813-5888]{David Wittman}\affiliation{Department of Physics \& Astronomy, University of California, Davis, CA 95616, USA}
\author[0000-0002-7701-9215]{Dariush Imani}\affiliation{Department of Physics \& Astronomy, University of California, Davis, CA 95616, USA}
\author{Rutger Hartmann Olden}\affiliation{Department of Physics \& Astronomy, University of California, Davis, CA 95616, USA}
\author[0000-0003-2632-572X]{Nathan Golovich}\affiliation{Lawrence
  Livermore National Laboratory, 7000 East Avenue, Livermore, CA
  94550, USA}

\correspondingauthor{David Wittman}
\email{dwittman@physics.ucdavis.edu}

\begin{abstract} 
  Studies of star formation in various galaxy cluster mergers have
  reached apparently contradictory conclusions regarding whether
  mergers stimulate star formation, quench it, or have no effect.
  Because the mergers studied span a range of time since pericenter
  (TSP), it is possible that the apparent effect on star formation is
  a function of TSP. We use a sample of 12 bimodal mergers to assess
  the star formation as a function of TSP.  We measure the equivalent
  width of the H-alpha emission line in ${\sim}100$ member galaxies in
  each merger, classify galaxies as emitters or non-emitters, and then
  classify emitters as star-forming galaxies (SFG) or active galactic
  nucleus (AGN) based on the
  [NII] $\lambda6583$ line.  We quantify the distribution of SFG and
  AGN relative to non-emitters along the spatial axis defined by the subcluster
  separation. The SFG and AGN fractions vary from merger to merger,
  but show no trend with TSP.  The spatial distribution of SFG
  is consistent with that of non-emitters in eight mergers, but show
  significant avoidance of the system center in the remaining four
  mergers, including the three with the lowest TSP.  If there is a
  connection between star formation activity and TSP, probing it
  further will require more precise TSP estimates and more mergers
  with TSP in the range of 0--400 Myr.
  \end{abstract}

\section{Introduction} \label{sec:intro}


A merger of galaxy clusters provides an important astrophysical
laboratory for probing the behavior of dark matter and the hot gas in
the intracluster medium; see \citet{Molnar16review} for a review.
Studies conflict on whether mergers stimulate star formation
\citep{MillerOwen2003, MaEbelingMarshall2010,Sobral2015, Stroe2017},
suppress it \citep{Mansheim2017hiz}, or have no substantial effect
\citep{Chung2010}. These studies typically focus on a single merger or
two, with some comparison to field galaxies at comparable redshift.
Hence, the disparate results are potentially consistent with mergers
having a variable effect depending on dynamical parameters such as
pericenter speed or the elapsed time between pericenter and the
observed state, or time since pericenter (TSP).  If star formation is
stimulated promptly then later quenched, for example, observations of
various mergers will naturally lead to differing conclusions.


Testing this hypothesis requires a sample of mergers for which TSP can
be inferred.  The classic ``timing argument'' \citet{Kahn1959} uses
equations of motion for two point masses on radial trajectories in an
expanding universe.  Further assumptions on the viewing angle allows
one to compare this model to observables: the projected subcluster
separation \dproj\ and the line-of-sight component of the relative
velocity, \dv. \citet{Dawson2012} updated this idealized model by
using Navarro-Frenk-White \citep{NFW97} mass profiles rather than
point masses, and by marginalizing over possible viewing angles, while
keeping in place the assumption of zero pericenter distance.

\citet{Analogs2018} and \citet{Wittman19analogs} developed an approach
that naturally incorporates a cosmologically motivated range of
pericenter distances, as well as effects of substructure and
surrounding structure. They identified post-pericenter halo pairs in
cosmological n-body simulations and placed mock observers to compute
the joint likelihood of viewing angle and halo pair, given the
observables \dproj\ and \dv, as well as the subcluster masses if
known.  The joint likelihood can be marginalized over viewing angle to
obtain a likelihood for each simulated halo pair to serve as an analog
for the observed system.  Furthermore, each halo pair is tagged with a
known TSP, obtained from prior snapshots of the cosmological
simulation.  This enables calculation of likelihoods for TSP as well
as other dynamical variables such as pericenter speed \vmax\ and angle
between velocity and separation vectors $\varphi$.
\citet{Wittman19analogs} noted that allowing for $\varphi\ne0$,
i.e. nonradial trajectories, removed a longstanding bias in TSP
inference.  In systems where any nonzero \dv\ is observed, the
assumption of radial trajectories immediately rules out a subcluster
separation vector in the plane of the sky. Real systems do have some
tangential velocity component, so plane-of-sky separation
vectors---the most likely configuration {\it a priori}---are falsely
ruled out under the radial trajectory assumption; this leads to a bias
toward larger 3-D separations, hence larger TSP.

Merging cluster samples have grown in size and quality as well.  \cite
[hereafter G19a]{MCCsampledata} conducted an optical and radial
velocity survey of 29 radio-selected mergers, and \cite[][hereafter
G19b]{MCCsampleanalysis} followed up with a panchromatic analysis
including X-ray and radio data. Their subclustering analysis, based on
${\approx}100$-200 spectroscopically confirmed member galaxies per
system, revealed some of these systems to be multimodal, hence not
amenable to modeling with halo pairs. Still, 15 systems were
identified as bimodal, with the subclustering analysis providing
best-fit central coordinates and velocities for each subcluster. This
enables consistent TSP and \vmax\ inferences for each of an ensemble
of cluster mergers.  Of these, ten systems at redshifts $z{\le}0.38$
have spectra that cover the \Ha\ line, an indicator of current star
formation. These spectra also cover the nearby [NII]
  $\lambda6583$ line, enabling us to separate those galaxies whose
\Ha\ emission is related to an active galactic nucleus (AGN) rather
than current star formation.

This paper revisits the spectroscopy of member galaxies in those
clusters, extracting equivalent widths of the \Ha\ and 
[NII] $\lambda6583$ lines to identify star forming galaxies
(SFG) and AGN.  We then examine whether the
fraction or spatial distribution of SFG and/or AGN differs from that
of other confirmed cluster members, and explore whether any such
differences may be a function of TSP or \vmax. The remainder of this
paper is organized as follows: \S\ref{sec:methods} describes our
methods; \S\ref{sec:results}, the results; and \S\ref{sec:discussion}
provides discussion and interpretation.

\section{Data and Methods} \label{sec:methods}


We focus on \Ha\ emission because it is a robust indicator of current
star formation---see \citet{Kennicutt2012} for a review---and because
most of the clusters in the G19a spectroscopic survey were targeted
over a wavelength range that includes \Ha.  Although the exact
wavelength coverage varies from slit to slit in the G19a survey,
nearly all slits that probe \Ha\ also probe [NII] $\lambda6583$,
allowing for discrimination between SFG and AGN as described below.

\textit{Cluster selection and geometry.}  We start with the 29
clusters in the G19a sample and select those for which G19a took
spectra that went red enough to capture \Ha\ for cluster members. This
excludes clusters beyond $z{\approx}0.38$; PLCKESZ G287.0+32.9 at
$z{=}0.38$ is the highest-redshift cluster examined in this work.
Next, following the subclustering results of G19b we selected those
systems that were determined to be bimodal.  This yields the 12
merging systems listed in Table~\ref{tab-clust}.


\begin{table*}
\centering 
\caption{Cluster parameters}
\begin{tabular}{lcrcccc}
& & &\multicolumn{2}{c}{Subcluster 1} & \multicolumn{2}{c}{Subcluster 2}\\
  Name & z & N\tablenotemark{a}& RA (J2000) & DEC & RA & DEC  \\\hline  
ZwCl 0008.8+5215 & 0.10 & 103 & 00:12:02.5  & 52:34:00.9 & 00:11:32.5 & 52:30:35.3\\
Abell 115 & 0.19 & 136 & 00:55:53.9  & 26:25:16.5 & 00:56:01.5 & 26:20:07.4\\
Abell 523 & 0.10 & 114 & 04:59:13.2  & 08:49:45.0 & 04:59:05.2 & 08:43:37.5\\
Abell 3411 & 0.16 & 215 & 08:41:51.7  & -17:27:20.6 & 08:42:05.9 & -17:34:01.8\\
RXC J1053.7+5452 & 0.07 & 75 & 10:53:35.5  & 54:52:07.5 & 10:54:08.5 & 54:49:48.0\\
Abell 1240 & 0.19 & 86 & 11:23:36.9  & 43:09:08.6 & 11:23:37.0 & 43:03:13.7\\
PLCKESZ G287.0+32.9 & 0.38 & 111 & 11:50:49.7  & -28:04:32.6 & 11:51:02.9 & -28:12:27.4\\
Abell 1612 & 0.18 & 52 & 12:47:34.6  & -02:46:54.2 & 12:47:52.8 & -02:49:37.6\\
RXC J1314.4-2515 & 0.25 & 133 & 13:14:34.5  & -25:15:58.2 & 13:14:22.4 & -25:16:24.5\\
ZwCl 1856.8+6616 & 0.30 & 45 & 18:56:35.8  & 66:21:14.8 & 18:56:31.7 & 66:25:04.5\\
Abell 2443 & 0.11 & 145 & 22:26:01.1  & 17:23:48.1 & 22:26:09.7 & 17:20:47.7\\
CIZA J2242.8+5301 & 0.19 & 201 & 22:42:49.7  & 53:04:32.9 & 22:42:41.6 & 52:58:06.3\\
\end{tabular}
\tablenotetext{a}{Number of galaxies within $3\sigma_v$ of the
  systemic velocity.}
\label{tab-clust}\end{table*}

We are interested in investigating \Ha\ emission as a function of
position along the merger axis, so we use the subcluster separation
vector to define an $x$ coordinate on the sky. The point midway
between the subclusters is defined as $x=0$, and the subclusters
themselves define $x=\pm1$. We do not use the center of mass to define
$x=0$ because mass measurements are not available for each subcluster.
With this definition the sign of $x$ is arbitrary, and our later
analysis will account for that by taking the absolute value of $x$.

\textit{Slit targeting.} The G19a slit targeting procedure probed
cluster members with a wide range of properties.  Five of the twelve
clusters considered here were in the Sloan Digital Sky Survey (SDSS)
footprint at the time of G19a targeting. For these systems (Abell 115,
Abell 1240, Abell 1613, Abell 2443, and RXC J1053.7+5452) G19a used
SDSS photometric redshifts, creating a ``Sample 1'' catalog consisting
of galaxies within $\pm0.05(1 + z_{\rm cluster} )$ of the cluster
redshift, with a higher priority for brighter galaxies. The
DSIMULATOR\footnote{\url{http://www.ucolick.org/~phillips/deimos_ref/masks.html}}
slit targeting software filled the mask with as many Sample 1 targets
as possible while avoiding slit collisions, then filled in the
remainder of the mask with Sample 2 targets. For the systems with
photometric redshifts, Sample 2 consisted of other SDSS galaxies, with
a higher priority for brighter galaxies.

Targeting for the other seven systems used a red sequence method based
on photometry available at the time, including the POSS-II Digitized
Sky Survey \citep{DSS92} for ZwCl 1856.8+6616; Isaac Newton
Telescope/Wide Field Camera data from \citet{RvW2011photometry} for
Abell 523, Abell 3411, ZwCl 0008.8+5215, and two CIZA J2242.8+5301
masks; and Subaru/SuprimeCam photometry for Abell 1314, PLCKESZ
G287.0+32.9, and the remaining two CIZA J2242.8+5301 masks.  For these
systems, Sample 1 consisted of bright ($r <22.5$) red sequence
galaxies and Sample 2 consisted of faint ($22.5 < r <23.5$) red
sequence galaxies. However, there was still a need for filler slits
and these were filled with blue cloud galaxies.  Furthermore, where
galaxies with known redshift were not targeted to avoid duplication,
this tended to deplete red sequence galaxies from the sample.

\begin{figure}
  \centering
    \includegraphics[width=\columnwidth]{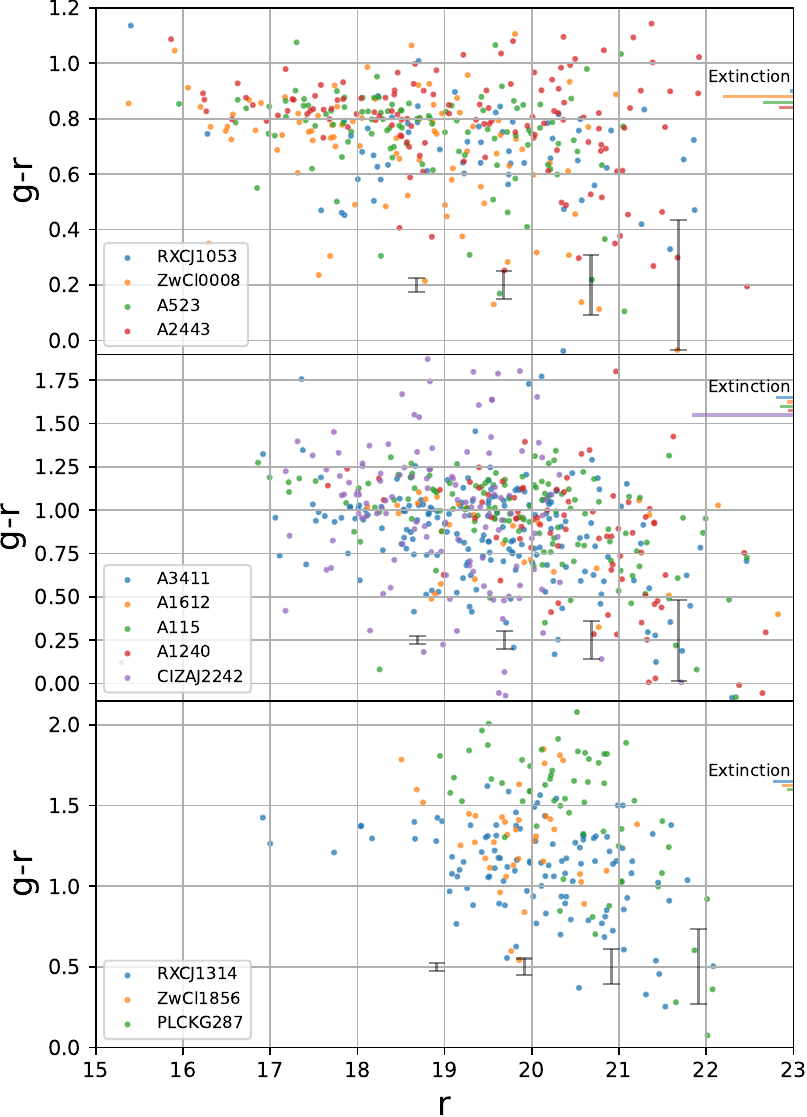}
    \caption{Pan-STARRS photometry of galaxies in the sample. The
      galaxies cover a wide range of luminosities and colors. Clusters
      are shown in order of increasing redshift, both within panels
      and across panels ($z{\approx}0.10,0.19$, and 0.30 from top to
      bottom). Colors have been dereddened; the extinction bars
      illustrate the widely varying extinction in $r$ band, which
      affects the depth to which obtaining a redshift is
      practical. Error bars show the typical color uncertainty at a
      given magnitude, taking into account the average extinction in a
      given panel; error bars smaller than the points have been
      omitted.}
    \label{fig-cmd}
\end{figure}

Furthermore, the long axis of the $16\times4$ arcmin DEIMOS subtends
1.3--5.0 Mpc depending on cluster redshift, with a median of 2.9
Mpc. This corresponds to 2--6 times the subcluster separation,
extending well beyond the central regions where red sequence galaxies
dominate.  Hence even where red sequence selection was used for
prioritization, the final sample includes a wide range of galaxy
colors. This is illustrated by Figure~\ref{fig-cmd}, which shows
Pan-STARRS \citep{PS1} photometry for the sake of uniformity,
specifically using $g-r$ colors because those two filters straddle the
4000 \AA\ break at these redshifts. Clusters are shown in order of
increasing redshift, with clusters of similar redshift grouped in each
panel ($z\approx 0.10,0.19$, and 0.30 from top to bottom).  The
photometry has been dereddened using values from the extinction
calculator provided by the NASA/IPAC Extragalactic
Database\footnote{\url{https://ned.ipac.caltech.edu/extinction_calculator}}
(NED).  The extinction in $r$ is shown as a bar for each cluster, to
help explain why some clusters seem to have shallower sampling than
others; before dereddening, all clusters were similar in terms of the
the faintest apparent magnitude at which a secure redshift was found.

In the low-redshift clusters, the galaxies sampled span a range of
nearly seven magnitudes. At the other extreme PLCKESZG287.0+32.9 is
sampled over a more limited three-magnitude range due to its higher
redshift ($z{=}0.38$), and CIZA J2242.8+5301 is sampled over a
similarly limited range due to heavy extinction. \textbf{The brightest
  galaxies sampled in each cluster have absolute $r$ magnitudes around
  $-22.5$.} Sampling along the color axis spans a range of one
magnitude or more in $g-r$.  The red sequence clearly moves up in
$g-r$ with redshift, and even the dereddened CIZA J2242.8+5301 is
consistent with this trend. However the scatter in CIZA J2242.8+5301
is large after dereddening due to spatially varying extinction not
modeled here. Setting that cluster aside, the range of colors sampled
is still quite large.  There is no clear correlation with the
targeting scheme; for example in the top panel the two clusters with
nominal red sequence selection (Abell 523 and ZwCl 0008.8+5215)
clearly sample the blue cloud better than one of the two clusters with
photometric redshift selection (RXC J1053.7+5452) and at least as well
as the other (Abell 2443). This illustrates how multiple factors (a
large spatial footprint, the need to avoid slit collisions in dense
regions, and so on) can diffuse the focus away from the red sequence
even in those clusters with nominal red sequence selection.

\textit{Data description.} We provide a brief summary and refer
readers to G19a for more details. Spectra were taken with the DEIMOS
multi-object spectrograph \citep{Faber2003} on the Keck II telescope
at the W. M. Keck Observatory. The the 1200 line mm$^{-1}$ grating was
used with $1^{\prime\prime}$ slits, yielding a pixel scale of 0.33
\AA\ pixel$^{-1}$, a resolution of $\sim 1$ \AA\ or 50 km s$^{-1}$,
and wavelength coverage spanning $\approx 2600$ \AA.  For these
lower-redshift clusters, the grating was tilted to ensure that \Ha\
was captured, e.g. spanning 5400--8000 \AA\ at a typical redshift of
0.2.  Within a cluster, the wavelength coverage varied by $\pm 400$
\AA\ depending on slit location.  The DEEP2 versions of the spec2d and
spec1d packages \citep{Deep2:2012,Deep2:2013} were used to remove
instrumental artifacts, calibrate the wavelengths, and extract 1-D
spectra.

\textit{Equivalent width.} The spectra were not photometrically
calibrated and the slits capture only the central part of each galaxy,
hence we do not attempt to extract \Ha\ luminosities. We rely on the
equivalent width (EW), which compares the line to the surrounding
continuum.  This comparison has the additional virtue of being robust
against dust in the emitting galaxy.

Using the galaxy redshifts determined by G19a, as well as the systemic
redshifts and velocity dispersions $\sigma_v$ from G19b, we selected
cluster members as those within $3\sigma_v$ of the systemic velocity.
We then created a nonparametric model of the continuum using an
iterative smoothing procedure. Ideally, the smoothing is done after
removing emission lines and other outliers such as bad pixels or
residuals from subtracted sky lines. Our first pass removes the most
intense 2.5\% of pixels and the least intense 1.25\% of pixels, then
smooths using a boxcar kernel 400 pixels (132 \AA) wide.  These
parameters were chosen for their robustness over a range of galaxy
spectral types and signal-to-noise levels.  The rms residuals from
this first-pass model are then calculated, and outliers are defined as
pixels whose first-pass residuals exceeed three times this rms
residual.  The second pass smooths the original data with the same
boxcar kernel, but now excluding outliers according to the above
criterion.

The redshift of Abell 3411 places the \Ha\ line in the 7600-7650 \AA\
telluric absorption feature.  The procedure described above was able
to distinguish emitters and non-emitters, but often ascribed positive
EW to the non-emitters, thus confusing \Ha\ absorption with telluric
absorption. To avoid any bias in the EW measurements for this cluster
we created an atmospheric transmission model using high
signal-to-noise galaxies with no evident \Ha\ features.  We then
divided all Abell 3411 spectra by this model before running the EW
extraction procedure described above. With this correction in place,
Abell 3411 shows no more tendency to \Ha\ absorption than any other
cluster. This correction had negligible effect on the results
described below, as emitters were already correctly identified.

We assess the uncertainty in the EW both empirically and with a noise
model.  The empirical method builds on the fact that the DEEP2
pipeline provides two different 1-D extractions of the underlying 2-D
spectral trace. The ``optimal'' extraction follows the algorithm of
\cite{Horne1986} with modifications described in \citet{Deep2:2013},
while the boxcar extraction uses a fixed width with a correction for
the fraction of light falling outside that fixed width. Each 1-D
extraction has an accompanying per-pixel uncertainty estimate. Both
extractions rely on the same 2-D spectrum so they are not
statistically independent; rather the extraction comparison is
designed to highlight systematic differences.  For example, the two
extractions differ visibly near the end of the observed wavelength
range, which becomes important for clusters where \Ha\ lies near the
end of that range. Different realizations of the continuum could
interact with our iterative continuum modeling procedure in
complicated ways, so we measure the EW from each extraction and
compare. For 68\% (95\%) of the galaxies, the two EW estimates were
within 0.19 (3.5) \AA. Looking more closely at galaxies where the two
EW estimates disagreed by more than 2 \AA\ (60 of the 746 galaxies, or
8\%) we found the vast majority of these to be strong emitters for
which this was a small fractional error.  We conclude that
uncertainties related to extraction do not impair our ability to
classify galaxies as emitters or nonemitters (we consider the few
ambiguous cases in more detail below).  We adopt the mean of the two
EW estimates as our final EW estimate.

Our noise model uses propagation of uncertainties from the per-pixel
uncertainties, assuming the analytically tractable case of modeling
the continuum as linear. In this limit, we find that the dominant
error for most slits arises from uncertainty in the pixels
representing \Ha, rather than in the continuum estimation. For 85\% of
slits, the uncertainty calculated this way is larger than the
difference between the two extractions, which is consistent with (i)
the dominant uncertainty arising from pixel statistics rather than
systematics and (ii) the two extractions not being statistically
independent. Hence we adopt the (usually more conservative)
uncertainties provided by the noise model. The galaxy coordinates, EW,
and uncertainties $\sigma_{\rm EW}$ are listed in Table~\ref{tab-EW}.

\startlongtable
\begin{deluxetable}{llrc}
  \tablecaption{\Ha\ equivalent width estimates}  \label{tab-EW}
  \tablecolumns{4}
  \tablehead{\colhead{RA} & \colhead{DEC} & \colhead{EW (\AA)} & \colhead{uncertainty (\AA)}}
  \startdata
  00:10:60.0 & 52:32:26.8 & 0.77 & 0.04\\
00:11:05.6 & 52:32:60.0 & 2.21 & 0.17\\
00:11:09.7 & 52:28:42.5 & -26.19 & 0.43\\
00:11:10.0 & 52:28:05.2 & 1.68 & 0.22\\
00:11:10.6 & 52:31:00.1 & 0.88 & 0.11\\

  \enddata \vskip5mm
\footnotesize{This is a sample of a long table, published as a
  machine-readable table (MRT) in \textit{Astronomical Journal.} For
  this arXiv version of the paper, we typeset the entire table at the end of the paper.}
\end{deluxetable}

\begin{figure}
  \centering
    \includegraphics[width=\columnwidth]{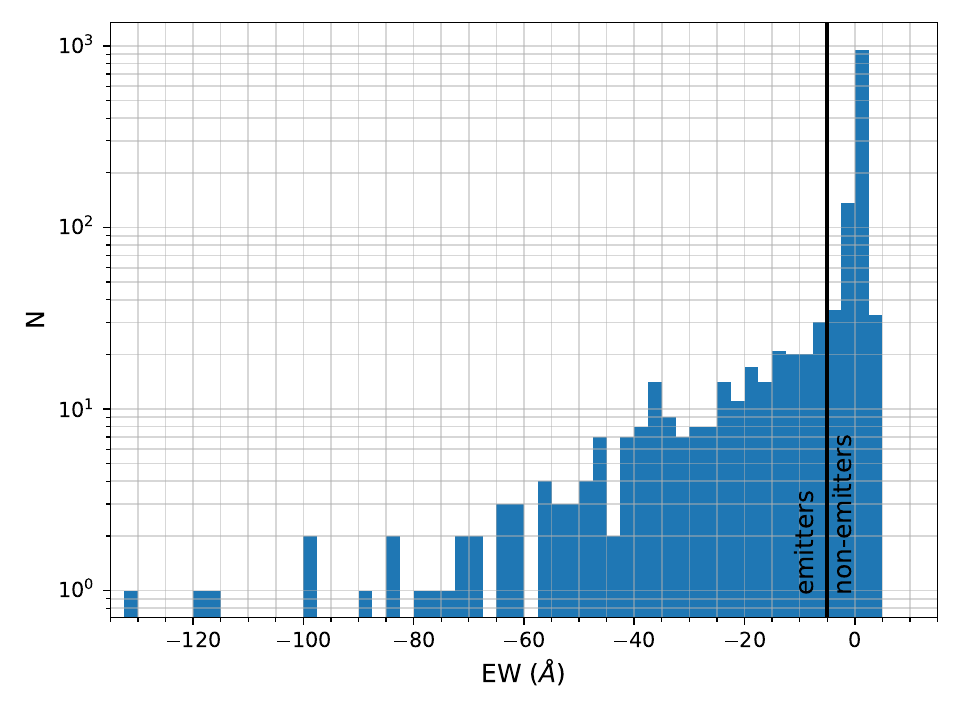}
    \caption{Histogram of \Ha\ EW measurements across all clusters and
    member galaxies.}
    \label{fig-EWhist}
\end{figure}

\textit{Emitter/nonemitter classification.} Figure~\ref{fig-EWhist}
shows a histogram of \Ha\ EW measurements across all clusters and member
galaxies. The distribution has a nearly symmetric peak around zero
with a long tail toward negative (emitting) values. The distinction
between the peak (whose width is determined by noise) and the tail of
emitters begins at an EW of -5 \AA. Therefore, we define three
categories:
\begin{itemize}
\item emitters: EW$ + 5\textrm{\AA} < -\sigma_{\textrm{EW}}$
\item nonemitters: EW$ +5\textrm{\AA} > \sigma_\textrm{EW}$
\item ambiguous: $|\textrm{EW}- 5\textrm{\AA}| <  \sigma_{\rm EW}$
\end{itemize}
where $\sigma_\textrm{EW}$ is the EW uncertainty of a given galaxy.
Of 1390 member galaxies, this yielded 1173 galaxies classified as
nonemitters, 240 as emitters, and 13 as ambiguous.

\textit{SFG/AGN classification.} \Ha\ emission may be due to
star formation or to AGN activity. Line ratios are useful in
distinguishing the two \citep{BPT1981,Carter01}. Based on Figure 10 of
\citet{Carter01} we adopt the categories:
\textbf{\begin{itemize}
\item  $\log_{10} \textrm{[NII]6583}/\textrm{H}\alpha < -0.3$: star-forming galaxy 
\item  $-0.3 < \log_{10} \textrm{[NII]6583}/\textrm{H}\alpha < -0.25$: ambiguous
\item $\log_{10} \textrm{[NII]6583}/\textrm{H}\alpha > -0.25$: AGN
\end{itemize}}
Of 1390 cluster members with spectra supporting \Ha\ and [NII]
measurements, we find 240 \Ha\ emitters, of which 197 are classified
as SFG and 37 as AGN\textbf{, with six ambiguous}.

\textit{Dynamical modeling.}  We use the method of
\citet{Wittman19analogs} to infer TSP, the maximum relative velocity
reached around the time of pericenter $v_{\rm max}$, and the viewing
angle $\theta$ (defined as 90$^\circ$ for a subcluster separation
vector in the plane of the sky) for each system. In this method, halo
pairs at the appropriate redshift are identified in the Big Multidark
Planck (BigMDPL) Simulation \citep{BigMDPL2016}. These pairs are then
tracked back in time to ensure that they have had one pericenter
passage by the observed redshift, because each observed system has
radio relics and X-ray morphology that require at least one pericenter
passage but are unlikely to persist after two such passages.  Each
halo pair meeting this requirement is referred to as an analog. Each
analog is then ``observed'' from 2627 lines of sight distributed over
a sphere, each line of sight defined by a colatitude $\theta$ and an
azimuth $\phi$, defined with respect to the subcluster separation
vector.  The inclusion of $\phi$ is an important detail relative to
more analytical methods which assume a velocity vector parallel to the
separation vector, or in other words a perfectly head-on collision.

The observables are simple: the projected separation between
subclusters; the relative line-of-sight velocity between subclusters
(as measured by member galaxy redshifts); and subcluster masses where
available. The likelihood of each (analog, $\theta$, $\phi$) tuple is
computed. The likelihood for $\theta$ is then computed by
marginalizing over analogs and over $\phi$.  (Other methods assuming a
relative velocity parallel to the separation vector
must---wrongly---place the separation vector away from the plane of
the sky if the observed component of the relative velocity vector is
nonzero; by marginalizing over $\phi$ this method correctly handles
that common situation.) The likelihood for TSP ($v_{\rm max}$) is
computed by assigning a TSP ($v_{\rm max}$) to each analog and
marginalizing over $\theta$ and $\phi$. 

For the observed values and uncertainties, we used the values from
G19b. This sample has substantial overlap with that of
\citet{Wittman19analogs}, but that work included only the G19b ``gold
sample'' which focused on systems with double radio
relics. Table~\ref{tab-dynmodel} presents the inferred parameter
values and confidence intervals (CI).

\begin{table*}
\centering 
\caption{Inferred merger dynamical parameters}
\label{tab-dynmodel}
\begin{tabular}{lrrrrc}
&\multicolumn{2}{c}{TSP}  &\multicolumn{2}{c}{$v_{\rm max}$}& view angle\\ 
Name&Myr&68\% CI & km/s&68\% CI& (deg, 68\% CI)\\\hline
Abell 115&346 &  215-451 &1978&1772-2172 &68-90\\
Abell 523& 510& 331-632 &2377&1995-2618 &58-86\\
Abell 1240&451&195-577 &2254&1979-2466 &64-88\\
Abell 1612&638& 447-782 &2544&2372-2908 &70-90\\
Abell 2443&277& 156-381 &1977&1521-2194 &72-90\\
Abell 3411&755& 476-971 &2544&2194-2808 &72-90\\
CIZA J2242.8+5301&608&378-937&2629&2403-2808 &68-90\\
PLCKESZ G287.0+32.9&519&341-812&2586&2263-3091 &68-88\\
RXC J1053.7+5452&270&125-351 &1954&1521-2203 &66-88\\
RXC J1314.4-2515&206&100-328 &2350&2018-2653 &26-62\\
ZwCl 0008.8+5215&802&516-897 &2360&2020-2461 &72-90\\
ZwCl 1856.8+6616&446&50-627 &2416&2036-2749 &66-90\\
\end{tabular}
\end{table*}
  
\section{Results}\label{sec:results}

\begin{figure}
  \centering 
    \includegraphics[width=\columnwidth]{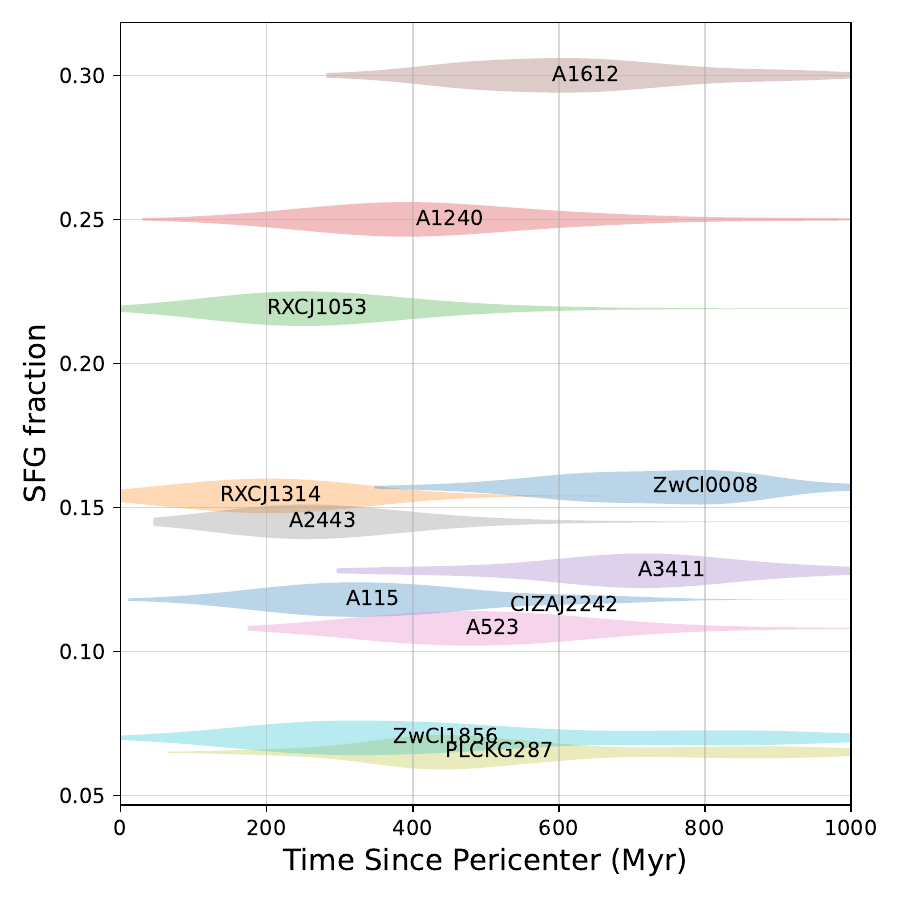}
    \caption{The SFG fraction versus time since pericenter. No trend
      is evident.}
    \label{fig-SFGfracTSP}
\end{figure}

\textit{Fraction of SFG and AGN.} Figure~\ref{fig-SFGfracTSP} plots
the fraction of members classified as SFG versus TSP, with each TSP
shown as a probability density function.  Although the TSP
distributions overlap, there is enough distinction between young and
old mergers to support visualization of a trend should it exist, but
no trend is evident.  The spectroscopic targeting was focused on
confirming likely cluster members based in part on color, so the
overall fraction of galaxies found to be emitters is likely to be
biased low; we defer discussion of this issue to
\S\ref{sec:discussion}.

\begin{figure}
  \centering 
    \includegraphics[width=\columnwidth]{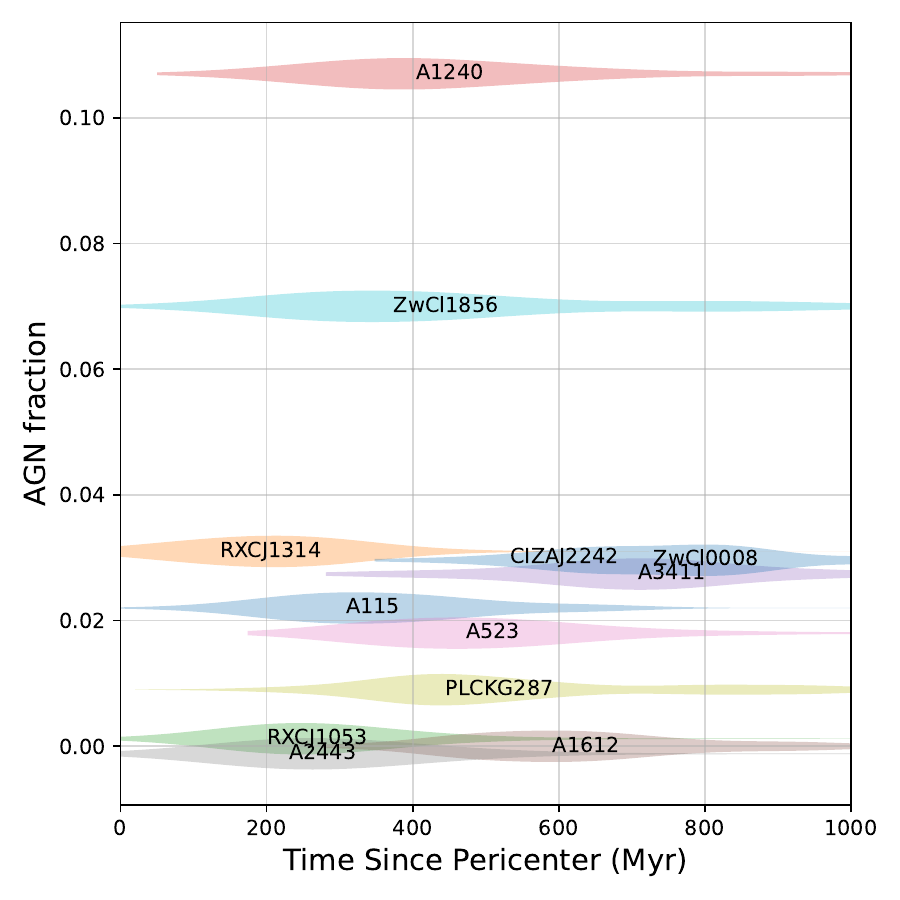}
    \caption{The AGN fraction  versus time since pericenter. No trend
      is evident.}
    \label{fig-AGNfracTSP}
\end{figure}

Figure~\ref{fig-AGNfracTSP} plots the fraction of members classified
as AGN versus TSP. Again, no trend is evident.  The reader may note
that Abell 1240 is high in both SFG fraction and AGN fraction. This
raises the question of whether clusters with high SFG fractions tend
to also have high AGN fractions. We find that this is not the case
($p=0.84$ for the null hypothesis of no correlation).

\begin{figure}
  \centering 
    \includegraphics[width=\columnwidth]{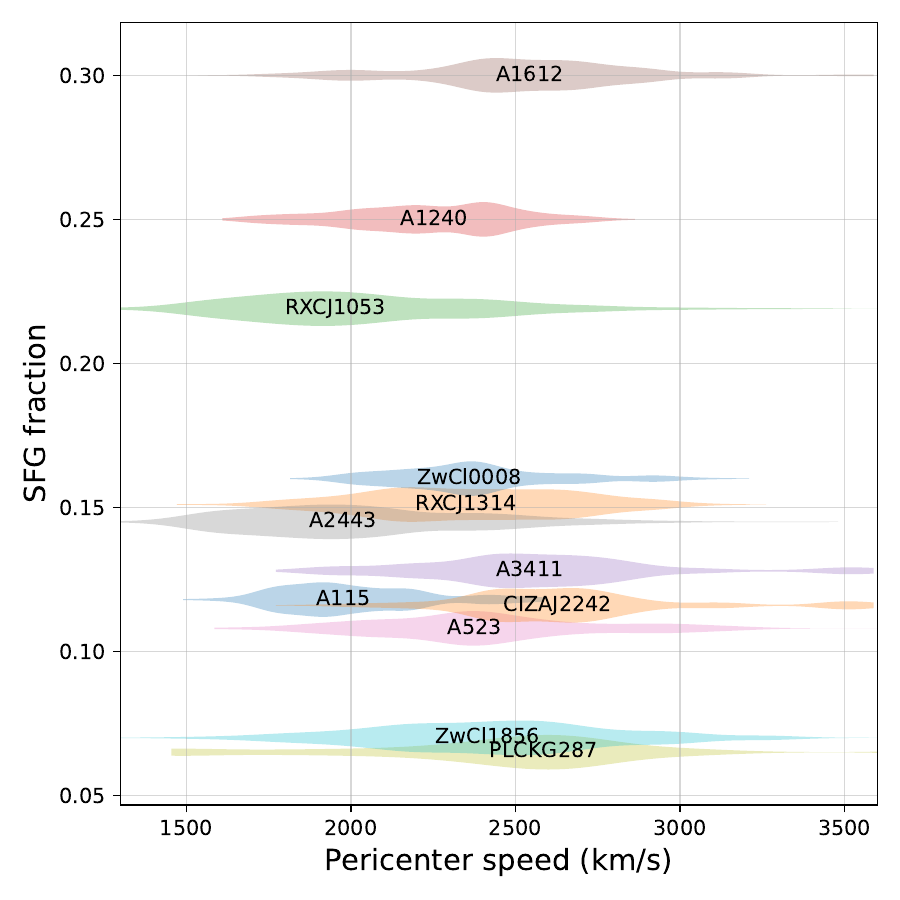}
    \caption{The SFG fraction shows no trend with pericenter
      speed. Small vertical shifts (0.003 or less) have been
      applied to some clusters for clarity.}
    \label{fig-SFGfracvmax}
\end{figure}

\begin{figure}
  \centering 
    \includegraphics[width=\columnwidth]{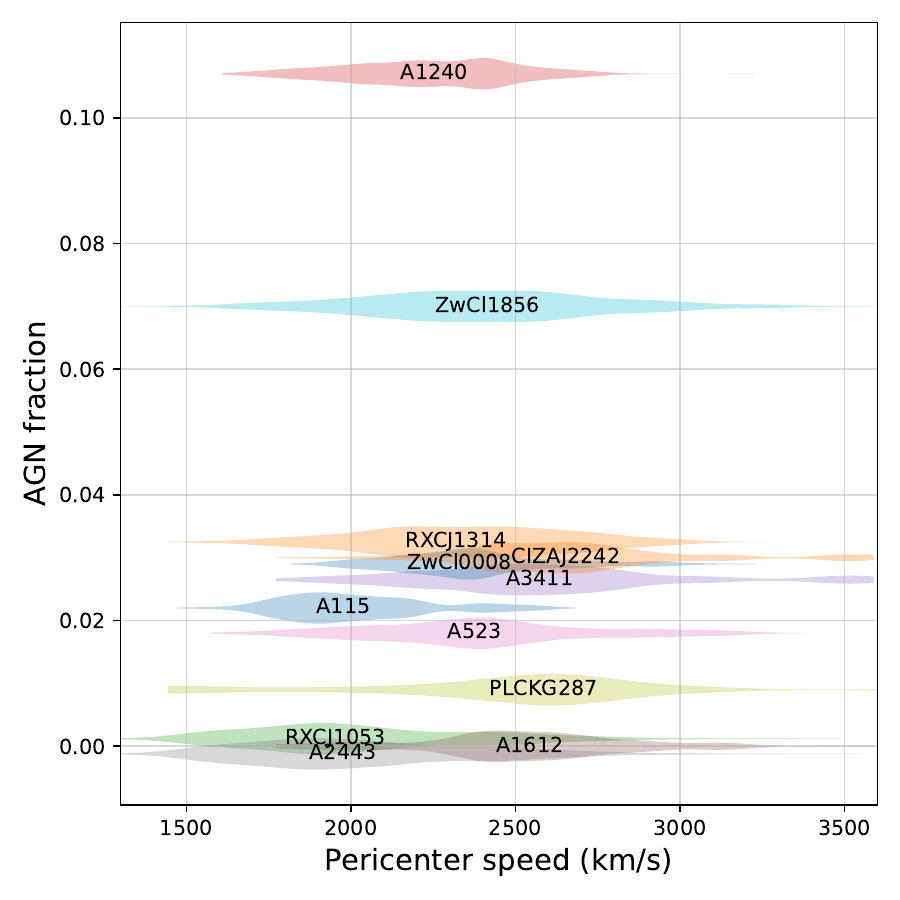}
    \caption{The AGN fraction shows no trend with pericenter
      speed. Small vertical shifts (0.002 or less) have been
      applied to some clusters for clarity.}
    \label{fig-AGNfracvmax}
\end{figure}

The pericenter speed estimates of various clusters overlap even more
than the TSP estimates, but for completeness we plot the SFG and
AGN fractions versus pericenter speed $v_{\rm max}$ in
Figures~\ref{fig-SFGfracvmax} and \ref{fig-AGNfracvmax}. There is no evident trend.

\begin{figure*}
  \centering 
  \includegraphics[width=\textwidth]{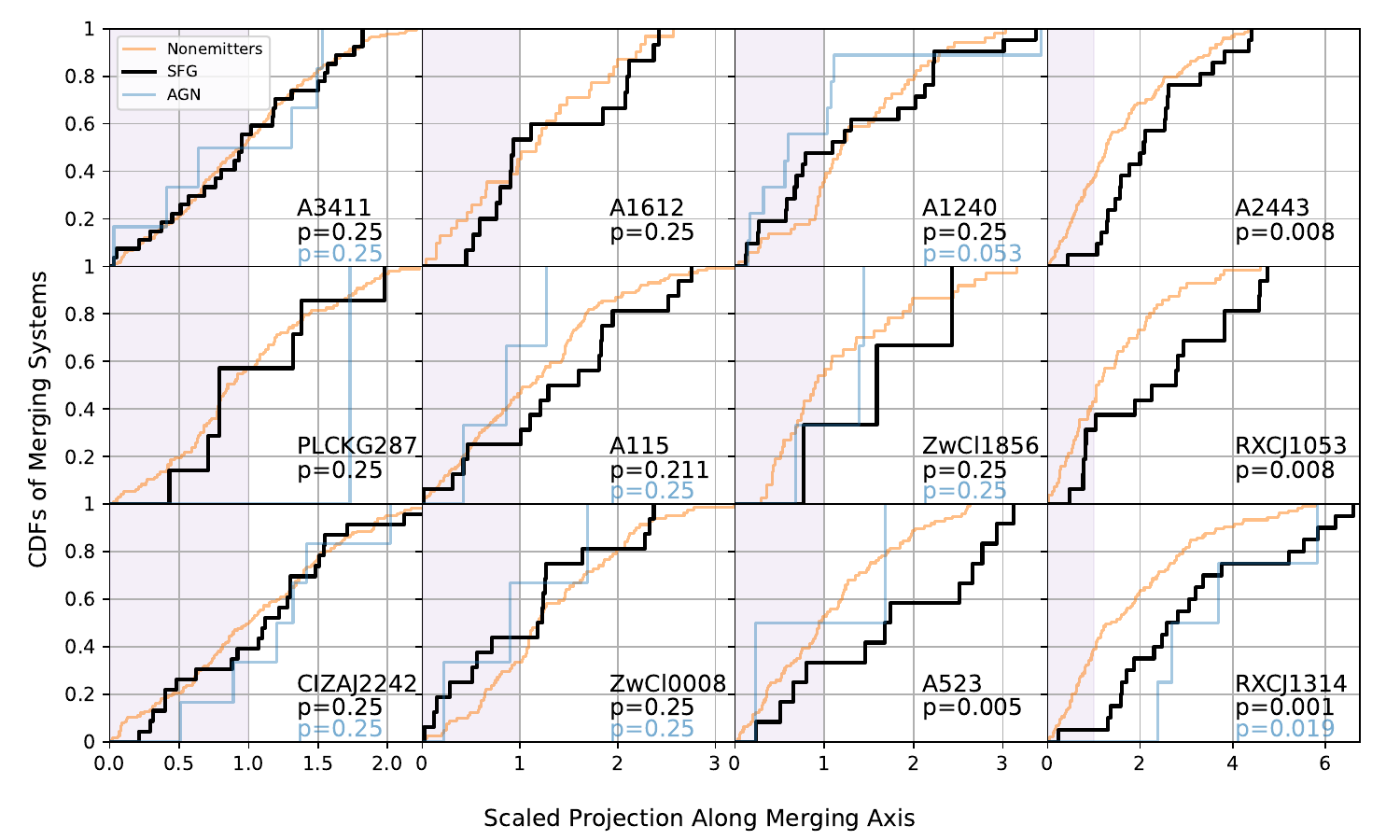}
  \caption{SFG and AGN vs. non-emitter spatial distribution for each
    of the merging systems. The distance along the merging axis is
    defined as zero at the center of the system and one at each of the
    subclusters, hence the shaded area represents the region between
    the subclusters. P-values are given for Anderson-Darling tests of
    SFG (black) and AGN (blue) consistency with the non-emitter
    distribution. For systems with two or fewer AGN, no AGN
    p-value is computed.}
    \label{fig-CDFs}
\end{figure*}

\textit{Spatial distribution of SFG and AGN.} Spectroscopic targeting
biases (such as slit collisions preventing full sampling of the
densest cluster regions) are likely to affect emitters and nonemitters
equally. Therefore we compare the spatial distribution of SFG
and AGN in a given system with the spatial distribution of
non-emitters in the same system.  Figure~\ref{fig-CDFs} shows the
cumulative distribution function (CDF) of the absolute value of $x$,
in other words the distance of each galaxy from the center of the
merging system, projected along the subcluster separation vector.
Hence, the shaded area represents the region between the subclusters'
centers.

In most systems, the distributions of emitters and non-emitters are
qualitatively similar. However, there are some notable variations. For
Abell 523, Abell 2443, RXC J1053.7+5452, and RXC J1314.4--2515 the
SFG CDF is consistently lower than the non-emitter CDF,
meaning that SFG tend to reside at larger projected distances
from the system center.  In three of these four systems, the paucity
of AGN prevents us from assessing whether AGN are distributed more
like the non-emitters or the SFG. But RXC J1314.4--2515 has sufficient
AGN to see that they are distributed similarly to the SFG rather
than the non-emitters.

It is difficult to find a counterexample: a system in which the SFG
and/or AGN are more centrally concentrated than the general
population. ZwCl 0008.8+5215 shows hints of both SFG and AGN being
more centrally concentrated, but this is not significant according to
the test described below. In Abell 1240 the AGN distribution appears
more centrally concentrated, but this has marginal significance
as described below.

To test the significance of any discrepancy between distributions we
employ the Anderson-Darling test which is a more sensitive version of
the classic Kolmogorov-Smirnov test for consistency between two
distributions \citep{2012msma.book.....F}.  We use the
\texttt{scipy.stats.anderson\_ksamp} routine, which returns a
truncated range of p-values between 0.001 (highly signficant) and 0.25
(no significance). These values are displayed in each panel of
Figure~\ref{fig-CDFs}. The four systems with far-flung SFG
listed above indeed have SFG distributions inconsistent
($p<0.01$) with their non-emitter distributions. Testing for
  consistency between AGN and non-emitter distributions, we find one
  system, Abell 1240, with a mild ($p=0.053$) tendency for centrally
  concentrated AGN, and one, RXC J1314.4--2515, with a stronger
  ($p=0.019$) tendency for far-flung AGN. In the latter system, the
  AGN distribution is consistent with the far-flung SFG distribution;
  the other three systems with far-flung SFG had too few AGN to
  support a meaningful test.

\textit{Relating SFG distribution to dynamical parameters.}  The two
clusters with the lowest TSP (RXC J1053.7+5452 and RXC J1314.4--2515)
have significantly far-flung SFG while the cluster with the highest
TSP (ZwCl 0008.8+5215) has SFG more centrally concentrated than
non-emitters (albeit not significantly so).  Hence, a natural
hypothesis is that the SFG distribution is affected by TSP.

\begin{figure}
  \centering 
    \includegraphics[width=\columnwidth]{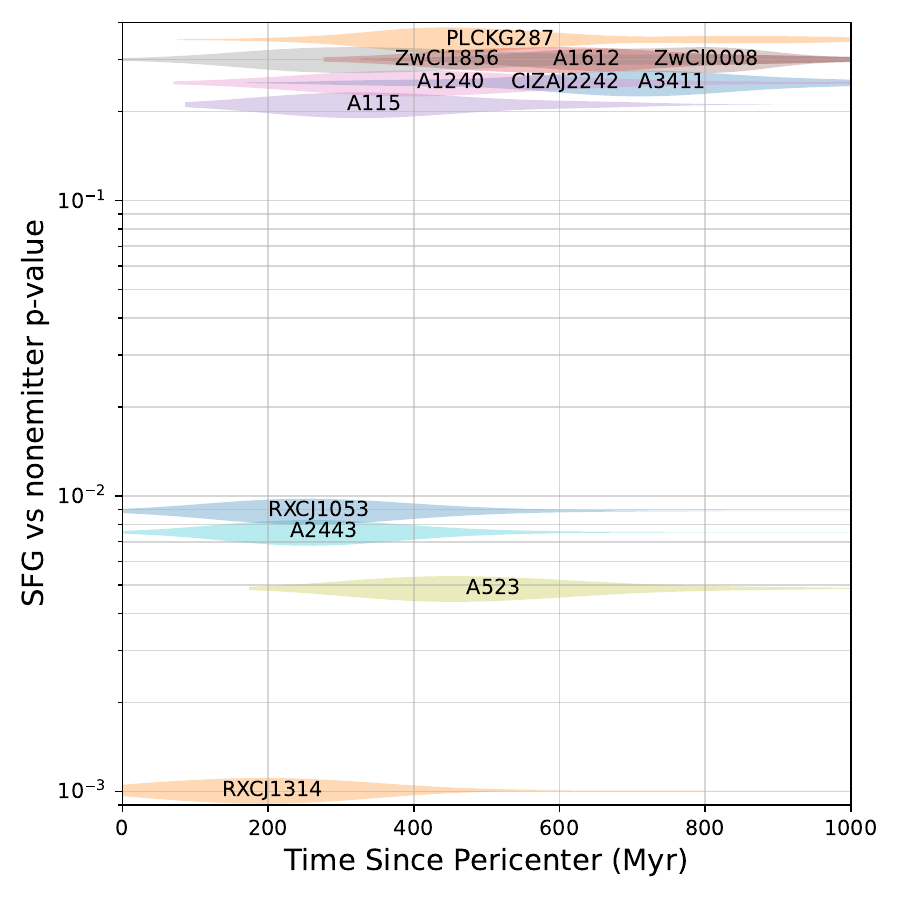}
    \caption{P-value (measuring agreement between the spatial
      distributions of SFG and of non-emitters) vs. TSP for each
      system. Lower p-values indicate that the SFG distribution along
      the merger axis differs from that of non-emitters. P-values are
      truncated to the interval [0.001,0.25]; the seven systems with
      entirely consistent distributions ($p{=}0.25$) are shown with
      slight vertical displacements for clarity. }
    \label{fig-SFGpvals}
\end{figure}

Figure~\ref{fig-SFGpvals} plots the SFG p-value versus TSP. The
vertical axis is defined such that systems with highly inconsistent
SFG and non-emitter distributions appear toward the bottom of
this plot. Of the four systems with standout SFG
distributions, three of them are also the three youngest systems. RXC
J1314 in particular is both the youngest system and the system with
the most significant discrepancy between emitter and non-emitter
distribution, at the p-value floor of 0.001. These examples
suggest that the SFG distribution may change with TSP, but
the case is far from clear: the fourth standout SFG
distribution, Abell 523, is middle aged.  We also tested the
SFG distribution as a function of radius from the center of
the system (projected on the sky) rather than along the merger
axis. The results show slightly weaker evidence for a trend with
projected radius.

\begin{figure}
  \centering 
    \includegraphics[width=\columnwidth]{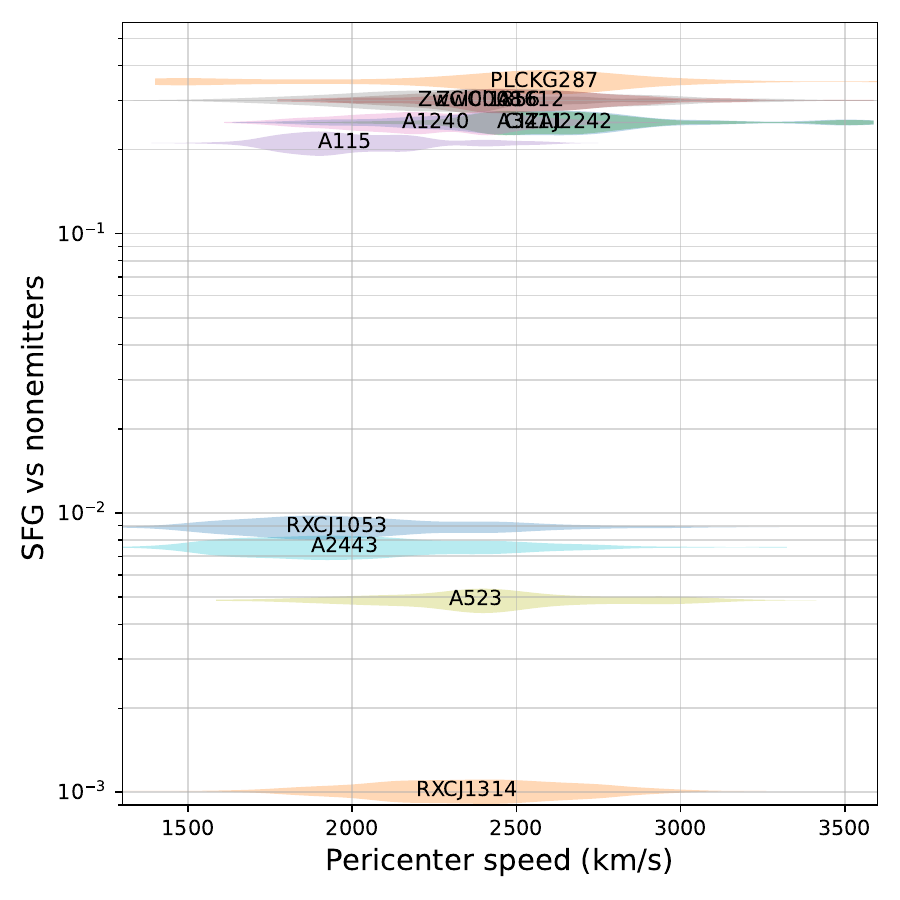}
    \caption{SFG spatial distribution vs. pericenter speed. See 
      Figure~\ref{fig-SFGpvals} for details.}
\label{fig-pval-vmax}
\end{figure}

Figure~\ref{fig-pval-vmax} plots the SFG p-value versus the
other dynamical parameter, pericenter speed. The four systems with
standout SFG distributions are evenly split between low and
moderate pericenter speeds.

\begin{figure}
  \centering 
    \includegraphics[width=\columnwidth]{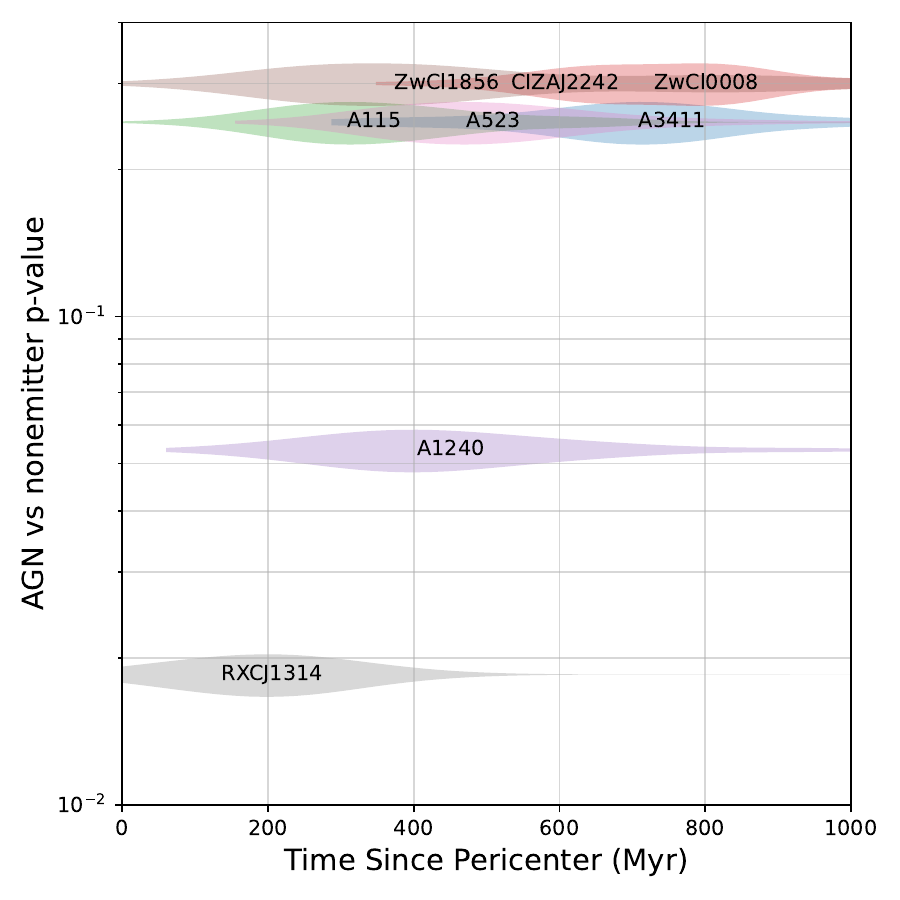}
    \caption{P-value (measuring agreement between the spatial
      distributions of AGN and of non-emitters) vs. TSP for each
      system. Lower p-values indicate that the AGN distribution along
      the merger axis differs from that of non-emitters. P-values are
      truncated to the interval [0.001,0.25]; the six systems with
      entirely consistent distributions ($p{=}0.25$) are shown with
      slight vertical displacements for clarity. Note that four
      systems have too few AGN to quantify their
      distribution along the merger axis.}
    \label{fig-AGNpvals}
\end{figure}

\textit{Relating AGN distribution to dynamical parameters.}
Figure~\ref{fig-AGNpvals} probes the relationship between TSP and the
spatial distribution of AGN. The cluster with the lowest TSP (RXC
J1314.4--2515) also has the most discrepant AGN distribution. However,
other clusters with low TSP (Abell 2443 and RXC J1053.7+5452), which
like RXC J1314.4--2515 had far-flung SFG distributions, have too few
AGN to probe their AGN spatial distribution. Hence, any suggested
relation between AGN spatial distribution and TSP rests largely on the
single example of RXC J1314.4--2515.

\begin{figure}
  \centering 
    \includegraphics[width=\columnwidth]{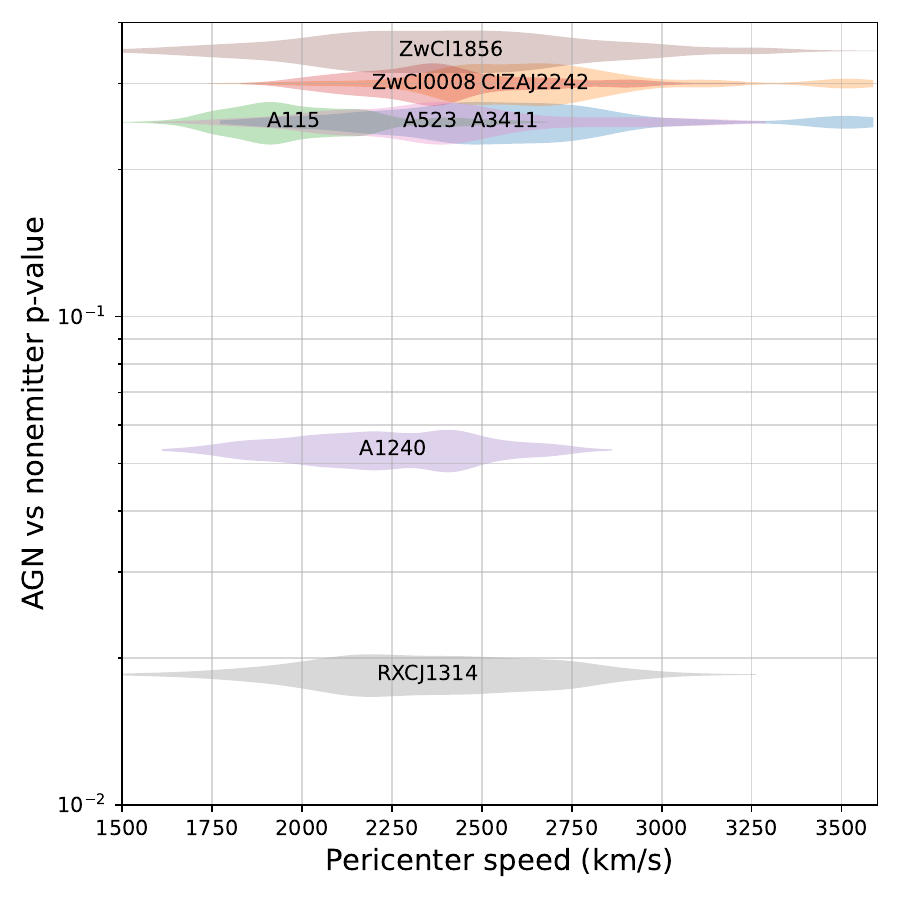}
    \caption{AGN spatial distribution vs. pericenter speed. See 
      Figure~\ref{fig-AGNpvals} for details.}
    \label{fig-AGNpval-vmax}
\end{figure}

Figure~\ref{fig-AGNpval-vmax} probes the relationship between AGN
spatial distribution and the other dynamical parameter, pericenter
speed. Here, there is not even a hint of a trend, as the two clusters
with measurably far-flung AGN distributions have unremarkable
pericenter speeds.

\section{Discussion}\label{sec:discussion}

\textit{Timescales.} \Ha\ emission is a indicator of very recent
\citep[past ${\sim}20$ Myr;][]{Kennicutt1998, Kennicutt2012} star
formation.  This is a much shorter timescale than the time since
pericenter in this sample, which ranges from 200--800 Myr.  Hence,
these data are not intended to probe any star formation that may have
been triggered and quenched earlier in the merger process.  Such
questions are better probed by using other spectral indicators
\citep{Mansheim2017DLS,Mansheim2017hiz}, combining morphology and
photometry with spectroscopy \citep{MaEbelingMarshall2010}, or
possibly stellar population synthesis modeling of member galaxies.
The narrow time window for \Ha\ emission is well suited to a
complementary question: where are the merger effects, if any, being
felt \textit{now}? A cluster merger is not a single event but a long
process which may affect different regions at different times. In
particular, pericenter passage initiates the launch of a shock that,
even at ${\sim}3000$ km/s, still takes $300$ Myr or more to travel 1
Mpc into the cluster outskirts.  The 16 arcmin DEIMOS field,
corresponding to 3.1 Mpc diameter at the median redshift of our
sample, is well suited to probing both shocked and unshocked regions.
  
Another relevant timescale is the galaxy dispersion timescale.
Consider each subcluster as an isolated unit with member galaxies
having random speeds of order 1000 km/s along any given direction $x$
on the sky. As a concrete example, if star formation were triggered
only at the time and location of pericenter, when viewed 500 Myr later
the affected galaxies will be dispersed over a 1 Mpc diameter window
along the merger axis. This makes it difficult to reconstruct events
that happened around the time of pericenter, even if detailed and
reliable star formation histories were extracted from all member
galaxies.  The limited time window relating \Ha\ emission to star
formation is actually an advantage in this regard: \Ha\ is emitted
promptly at the time {\it and location} of star formation.  
   
\begin{figure}
  \centering 
    \includegraphics[width=\columnwidth]{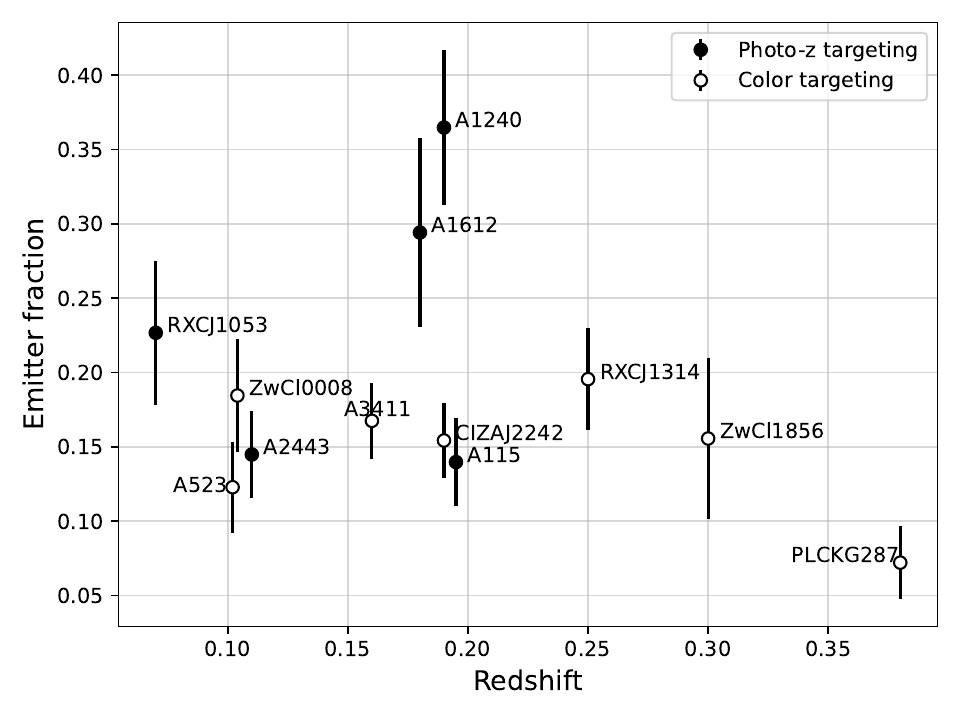}
    \caption{The emitter fraction shows no clear trend with redshift,
      but does appear to be lower with color selection, as expected if
      most slits are assigned to red-sequence galaxies. Uncertainties
      assume a binomial process operating independently in each
      cluster. }
    \label{fig-emfracz}
\end{figure}

\textit{Potential systematic and physical effects.} The fraction of
slits with SFG varies widely, from 0.065 (PLCKESZ
G287.0+32.9) to 0.30 (Abell 1612), but there is no
correlation between SFG fraction and TSP or $v_{\rm
  max}$. Similar statements apply to the fraction of slits
  with AGN, and to the fraction of slits with \Ha\ emitters of any
  type.  This raises the question of what does control the emitter
fraction.  Slit targeting may play a role.  For seven systems, masks
were designed based on two-band photometry with priority placed on red
sequence galaxies, while the other five systems were based on SDSS
photometry and photometric redshifts, which increased the likelihood
of targeting member galaxies outside the red sequence.
Figure~\ref{fig-emfracz} plots the emitter fraction using different
symbol types for the two types of slit targeting.  Indeed, the three
highest emitter fractions are in systems with photometric redshift
targeting.  Even in red-sequence masks, however, one-third to perhaps
one-half the slits were filler slits of bluer galaxies with magnitudes
appropriate to cluster members. This explains the substantial overlap
in emitter fraction between the two sets of targets in
Figure~\ref{fig-emfracz}.

Nevertheless, one can ask whether there are trends within each subset.
Figure~\ref{fig-emfracz} shows no obvious redshift trend within either
subset. This is perhaps surprising: given the fixed DEIMOS field of
view (about 16x5 arcmin, with the long dimension placed along the
subcluster separation vector), higher-redshift systems are probed to
larger physical radius where more emitters may be expected.

Next, we consider the spatial distribution of SFG and AGN
compared to non-emitters in the same system, focusing on SFG
  due to the paucity of AGN. Figure~\ref{fig-CDFs} shows clear
variations from system to system.  Figure~\ref{fig-SFGpvals} suggests
a potential association with TSP, but the evidence is weak.  Note that
slit targeting effects seem to play no role here: the four systems
with discrepant emitter distributions are evenly split between color
and photometric redshift targeting methods.  We also checked for an
association with redshift, and found none. It is possible that other
merger parameters, such as pericenter speed, have an effect, but the
uncertainties on the pericenter speed are currently too large to
differentiate clearly between slow and fast mergers.

What other effects could play a role in the SFG distribution?
Figure~\ref{fig-CDFs} does show that the systems with far-flung
SFG distributions tend to be those in which the spectroscopic
footprint, {\it in units of the projected subcluster separation}, is
larger.  The physical field size effect discussed above is not at play
here, again because that would cause a clear redshift trend that is
not seen here.  In fact the highest-redshift cluster, PLCKESZ
G287.0+32.9, is an illustrative counterexample: it is probed out to a
large physical radius, but a fairly small radius in units of subcluster
separations, and it shows no SFG/non-emitter discrepancy.

Perhaps the subcluster separation sets a spatial scale that affects
the SFG distribution.  One possible mechanism is the shock
generated at pericenter passage. The shock outruns the subclusters but
not by a large factor, at least in the first ${\sim}200$ Myr
\citep{Springel2007}, so the subcluster separation may be providing a
rough scale for the shock separation.  However, this does not explain
why, in three of the four systems with standout SFG distributions,
the SFG and non-emitter distributions start diverging {\it within}
one subcluster separation.

\textit{Early vs late type galaxies.} Merger effects on galaxies
likely depend on galaxy type. While a full exploration is beyond the
scope of this paper, we address one potential systematic effect here:
does the selection of many red sequence galaxies (thought to be
relatively immune to star formation triggers) lead predictably to a
null result?  We note that most mergers examined here \textit{have no
  dearth of SFG projected near the subcluster cores} ($x=1$ in our
coordinate system), where red sequence galaxies tend to congregate.
The fact that SFG\footnote{The morphological types of these emitters
  may be a topic for future work; we simply note here that projection
  effects and slit targeting effects allow the possibility that these
  emitters are not necessarily early type galaxies.}  are typically
found in these regions suggests that the G19a slit selection did not
result in low sensitivity to star formation activity.  The question is
rather: what is different about the four systems with star formation
activity located further out? As noted above, slit targeting
differences are not an explanatory variable.  Figure~\ref{fig-cmd}
suggests that variations in color sampling and depth are also not
explanatory. 

\textit{Relation to shock.} We checked the four systems with far-flung
star formation activity against the G19b diagrams of these systems,
illustrating radio relic positions (assumed to be tracing a shock)
relative to the subclusters, and we found no commonality. Three of the
systems have single relics and one (RXC J1314.4-2515) has double
relics, a ratio consistent with the G19b sample as a whole.  Two of
the systems (Abell 523 and Abell 2443) have shocks projected more or
less at the subcluster position ($x=1$ in our coordinate system),
while the other two have relics projected much further out
($x\approx 4$ for RXC J1053.7+5452 and RXC J1314.4-2515, with the
latter cluster also having a second relic projected at
$x \approx2.5$).  RXC J1314.4-2515 is noted as having substantial
line-of-sight velocity difference between the subclusters; the others
are not, a ratio consistent with the G19b sample as a whole.

\textit{Relation to other results.}  \citet{Sobral2015} proposed a
link between the shock and \Ha\ emitters in CIZA J2242.8+5301. They
used narrow-band imaging to select candidate emitters and confirmed 83
\Ha\ emitters spectroscopically, most with the AutoFib2+WYFFOS (AF2)
instrument on the William Herschel Telescope. The DEIMOS spectra used
here serendipitously provided confirmation for 26 of their emitter
candidates. The AF2 instrument covers a much larger field, with an
unvignetted field of 20$^\prime$ radius, and the fibers were targeted
specifically at narrow-band emitter candidates. Hence, the
\citet{Sobral2015} data set provides a complementary view of CIZA
J2242.8+5301 despite some overlap in spectroscopic data.  They propose
that \Ha\ emitters ``are found preferably near the shock fronts'' but
this is not evident in our (more limited) view of the same cluster.
If emitters had a preferred location (relative to non-emitters) their
CDFs in Figure~\ref{fig-CDFs} would look more like sigmoid functions,
rising sharply at the preferred location.  The most plausible example
of such a rise in Figure~\ref{fig-CDFs} is for ZwCl 0008.8+5215,
indeed just outside the subcluster separation.  However, this rise
represents only five SFG.

One caveat in our analysis is that we have not adopted a
center-of-mass (CM) coordinate system because the subcluster masses
are not all known.  It is possible that a CM analysis would allow some
pattern to emerge. However, we note that CIZA J2242.8+5301 has
subclusters of roughly equal mass \citep{Jee15CIZA} so a CM analysis
would not substantially change our result for that cluster.

Based on narrow-band imaging, \citet{Stroe15CIZA} found that CIZA
J2242.8+5301 ``has an \Ha\ emitter density ${>}20$ times that of blank
fields'' and \citet{Stroe17Survey} found that merging clusters have
double the \Ha\ emitters of relaxed clusters. This is combatible with
our results, as we do not compare to non-merging clusters or the
field, merely to non-emitters in the same merging cluster.
\citet{Stroe17Survey} also found that $\phi^*$, the characteristic
density of the \Ha\ luminosity function, drops from cluster core to
outskirts. This is also compatible with our results, because
non-emitter density also drops from cluster core to outskirts. In most
of our systems, these two densities drop at a similar rate, but in
some systems (Abell 523, Abell 2443, RXC J1053.7+5452, and RXC
  J1314.4--2515) the density of non-emitters actually drops faster
than that of \Ha\ emitters, resulting in more far-flung
  emitters than the general population targeted for slits. After
  classifying emitters as SFG or AGN, we find that this statement
  remains true for SFG in all four of those systems, \textit{and} for
  AGN in the one system (RXC J1314.4--2515) with sufficient AGN to
  make a statement.

In conclusion, these data are suggestive that in systems viewed soon
(${\approx}200$ Myr) after pericenter, \Ha\ emitting galaxies
are less centrally concentrated than non-emitting member
  galaxies in our spectroscopic sample. However, this is based
on three merging systems, and a fourth system with larger TSP also
exhibits this behavior. Probing this potential link further will
require more precise TSP estimates and more mergers with TSP in the
range of 0--400 Myr.  The \Ha\ emitters we find are 
  overwhelmingly star-forming galaxies, such that AGN distributions
  are often unconstrained by these data.   Nevertheless, it is worth
  noting that in the one low-TSP system where the AGN distribution can
  be constrained, it agrees with the SFG distribution rather than the
  non-emitter distribution.

\begin{acknowledgments}
  We thank the anonymous referee for several insightful suggestions.
  We thank Rodrigo Alves Stancioli for numerous suggestions and
  corrections.  The data presented herein were obtained at the
  W.M. Keck Observatory, which is operated as a scientific partnership
  among the California Institute of Technology, the University of
  California and the National Aeronautics and Space
  Administration. The Observatory was made possible by the generous
  financial support of the W.M. Keck Foundation.  Funding for the
  DEEP2/DEIMOS pipelines has been provided by NSF grant
  AST-0071048. The DEIMOS spectrograph was funded by grants from CARA
  (Keck Observatory) and UCO/Lick Observatory, a NSF Facilities and
  Infrastructure grant (ARI92-14621), the Center for Particle
  Astrophysics, and by gifts from Sun Microsystems and the Quantum
  Corporation.  Part of this work was performed under the auspices of
  the U.S. Department of Energy by Lawrence Livermore National
  Laboratory under Contract DE-AC52-07NA27344. This research has made
  use of NASA’s Astrophysics Data System. This research has made use
  of the NASA/IPAC Extragalactic Database (NED), which is funded by
  the National Aeronautics and Space Administration and operated by
  the California Institute of Technology.

\end{acknowledgments}\facility{Keck:II}


\bibliography{ms}{}

\begin{thebibliography}{}
\expandafter\ifx\csname natexlab\endcsname\relax\def\natexlab#1{#1}\fi
\providecommand{\url}[1]{\href{#1}{#1}}
\providecommand{\dodoi}[1]{doi:~\href{http://doi.org/#1}{\nolinkurl{#1}}}
\providecommand{\doeprint}[1]{\href{http://ascl.net/#1}{\nolinkurl{http://ascl.net/#1}}}
\providecommand{\doarXiv}[1]{\href{https://arxiv.org/abs/#1}{\nolinkurl{https://arxiv.org/abs/#1}}}

\bibitem[{{Baldwin} {et~al.}(1981){Baldwin}, {Phillips}, \&
  {Terlevich}}]{BPT1981}
{Baldwin}, J.~A., {Phillips}, M.~M., \& {Terlevich}, R. 1981, \pasp, 93, 5,
  \dodoi{10.1086/130766}

\bibitem[{{Carter} {et~al.}(2001){Carter}, {Fabricant}, {Geller}, {Kurtz}, \&
  {McLean}}]{Carter01}
{Carter}, B.~J., {Fabricant}, D.~G., {Geller}, M.~J., {Kurtz}, M.~J., \&
  {McLean}, B. 2001, \apj, 559, 606, \dodoi{10.1086/322349}

\bibitem[{{Chambers} {et~al.}(2016){Chambers}, {Magnier}, {Metcalfe},
  {Flewelling}, {Huber}, {Waters}, {Denneau}, {Draper}, {Farrow}, {Finkbeiner},
  {Holmberg}, {Koppenhoefer}, {Price}, {Rest}, {Saglia}, {Schlafly}, {Smartt},
  {Sweeney}, {Wainscoat}, {Burgett}, {Chastel}, {Grav}, {Heasley}, {Hodapp},
  {Jedicke}, {Kaiser}, {Kudritzki}, {Luppino}, {Lupton}, {Monet}, {Morgan},
  {Onaka}, {Shiao}, {Stubbs}, {Tonry}, {White}, {Ba{\~n}ados}, {Bell},
  {Bender}, {Bernard}, {Boegner}, {Boffi}, {Botticella}, {Calamida},
  {Casertano}, {Chen}, {Chen}, {Cole}, {Deacon}, {Frenk}, {Fitzsimmons},
  {Gezari}, {Gibbs}, {Goessl}, {Goggia}, {Gourgue}, {Goldman}, {Grant},
  {Grebel}, {Hambly}, {Hasinger}, {Heavens}, {Heckman}, {Henderson}, {Henning},
  {Holman}, {Hopp}, {Ip}, {Isani}, {Jackson}, {Keyes}, {Koekemoer}, {Kotak},
  {Le}, {Liska}, {Long}, {Lucey}, {Liu}, {Martin}, {Masci}, {McLean}, {Mindel},
  {Misra}, {Morganson}, {Murphy}, {Obaika}, {Narayan}, {Nieto-Santisteban},
  {Norberg}, {Peacock}, {Pier}, {Postman}, {Primak}, {Rae}, {Rai}, {Riess},
  {Riffeser}, {Rix}, {R{\"o}ser}, {Russel}, {Rutz}, {Schilbach}, {Schultz},
  {Scolnic}, {Strolger}, {Szalay}, {Seitz}, {Small}, {Smith}, {Soderblom},
  {Taylor}, {Thomson}, {Taylor}, {Thakar}, {Thiel}, {Thilker}, {Unger},
  {Urata}, {Valenti}, {Wagner}, {Walder}, {Walter}, {Watters}, {Werner},
  {Wood-Vasey}, \& {Wyse}}]{PS1}
{Chambers}, K.~C., {Magnier}, E.~A., {Metcalfe}, N., {et~al.} 2016, arXiv
  e-prints, arXiv:1612.05560, \dodoi{10.48550/arXiv.1612.05560}

\bibitem[{{Chung} {et~al.}(2010){Chung}, {Gonzalez}, {Clowe}, {Markevitch}, \&
  {Zaritsky}}]{Chung2010}
{Chung}, S.~M., {Gonzalez}, A.~H., {Clowe}, D., {Markevitch}, M., \&
  {Zaritsky}, D. 2010, \apj, 725, 1536, \dodoi{10.1088/0004-637X/725/2/1536}

\bibitem[{{Cooper} {et~al.}(2012){Cooper}, {Newman}, {Davis}, {Finkbeiner}, \&
  {Gerke}}]{Deep2:2012}
{Cooper}, M.~C., {Newman}, J.~A., {Davis}, M., {Finkbeiner}, D.~P., \& {Gerke},
  B.~F. 2012, Astrophysics Source Code Library

\bibitem[{{Dawson}(2013)}]{Dawson2012}
{Dawson}, W.~A. 2013, \apj, 772, 131, \dodoi{10.1088/0004-637X/772/2/131}

\bibitem[{{Djorgovski} {et~al.}(1992){Djorgovski}, {Lasker}, {Weir}, {Postman},
  {Reid}, \& {Laidler}}]{DSS92}
{Djorgovski}, S., {Lasker}, B.~M., {Weir}, W.~N., {et~al.} 1992, in American
  Astronomical Society Meeting Abstracts, Vol. 180, American Astronomical
  Society Meeting Abstracts \#180, 13.07

\bibitem[{{Faber} {et~al.}(2003){Faber}, {Phillips}, {Kibrick}, {Alcott},
  {Allen}, {Burrous}, {Cantrall}, {Clarke}, {Coil}, {Cowley}, {Davis}, {Deich},
  {Dietsch}, {Gilmore}, {Harper}, {Hilyard}, {Lewis}, {McVeigh}, {Newman},
  {Osborne}, {Schiavon}, {Stover}, {Tucker}, {Wallace}, {Wei}, {Wirth}, \&
  {Wright}}]{Faber2003}
{Faber}, S.~M., {Phillips}, A.~C., {Kibrick}, R.~I., {et~al.} 2003, in Society
  of Photo-Optical Instrumentation Engineers (SPIE) Conference Series, Vol.
  4841, Instrument Design and Performance for Optical/Infrared Ground-based
  Telescopes, ed. M.~{Iye} \& A.~F.~M. {Moorwood}, 1657--1669,
  \dodoi{10.1117/12.460346}

\bibitem[{{Feigelson} \& {Babu}(2012)}]{2012msma.book.....F}
{Feigelson}, E.~D., \& {Babu}, G.~J. 2012, {Modern Statistical Methods for
  Astronomy} (UK: Cambridge University Press)

\bibitem[{{Golovich} {et~al.}(2019{\natexlab{a}}){Golovich}, {Dawson},
  {Wittman}, {Jee}, {Benson}, {Lemaux}, {van Weeren}, {Andrade-Santos},
  {Sobral}, {de Gasperin}, {Br{\"u}ggen}, {Brada{\v{c}}}, {Finner}, \&
  {Peter}}]{MCCsampledata}
{Golovich}, N., {Dawson}, W.~A., {Wittman}, D.~M., {et~al.} 2019{\natexlab{a}},
  \apjs, 240, 39, \dodoi{10.3847/1538-4365/aaf88b}

\bibitem[{{Golovich} {et~al.}(2019{\natexlab{b}}){Golovich}, {Dawson},
  {Wittman}, {van Weeren}, {Andrade-Santos}, {Jee}, {Benson}, {de Gasperin},
  {Venturi}, {Bonafede}, {Sobral}, {Ogrean}, {Lemaux}, {Brada{\v{c}}},
  {Br{\"u}ggen}, \& {Peter}}]{MCCsampleanalysis}
---. 2019{\natexlab{b}}, \apj, 882, 69, \dodoi{10.3847/1538-4357/ab2f90}

\bibitem[{{Horne}(1986)}]{Horne1986}
{Horne}, K. 1986, \pasp, 98, 609, \dodoi{10.1086/131801}

\bibitem[{{Jee} {et~al.}(2015){Jee}, {Stroe}, {Dawson}, {Wittman}, {Hoekstra},
  {Br{\"u}ggen}, {R{\"o}ttgering}, {Sobral}, \& {van Weeren}}]{Jee15CIZA}
{Jee}, M.~J., {Stroe}, A., {Dawson}, W., {et~al.} 2015, \apj, 802, 46,
  \dodoi{10.1088/0004-637X/802/1/46}

\bibitem[{{Kahn} \& {Woltjer}(1959)}]{Kahn1959}
{Kahn}, F.~D., \& {Woltjer}, L. 1959, \apj, 130, 705, \dodoi{10.1086/146762}

\bibitem[{{Kennicutt}(1998)}]{Kennicutt1998}
{Kennicutt}, Robert~C., J. 1998, \apj, 498, 541, \dodoi{10.1086/305588}

\bibitem[{{Kennicutt} \& {Evans}(2012)}]{Kennicutt2012}
{Kennicutt}, R.~C., \& {Evans}, N.~J. 2012, \araa, 50, 531,
  \dodoi{10.1146/annurev-astro-081811-125610}

\bibitem[{{Klypin} {et~al.}(2016){Klypin}, {Yepes}, {Gottl{\"o}ber}, {Prada},
  \& {He{\ss}}}]{BigMDPL2016}
{Klypin}, A., {Yepes}, G., {Gottl{\"o}ber}, S., {Prada}, F., \& {He{\ss}}, S.
  2016, \mnras, 457, 4340, \dodoi{10.1093/mnras/stw248}

\bibitem[{{Ma} {et~al.}(2010){Ma}, {Ebeling}, {Marshall}, \&
  {Schrabback}}]{MaEbelingMarshall2010}
{Ma}, C.~J., {Ebeling}, H., {Marshall}, P., \& {Schrabback}, T. 2010, \mnras,
  406, 121, \dodoi{10.1111/j.1365-2966.2010.16673.x}

\bibitem[{{Mansheim} {et~al.}(2017{\natexlab{a}}){Mansheim}, {Lemaux},
  {Dawson}, {Lubin}, {Wittman}, \& {Schmidt}}]{Mansheim2017DLS}
{Mansheim}, A.~S., {Lemaux}, B.~C., {Dawson}, W.~A., {et~al.}
  2017{\natexlab{a}}, \apj, 834, 205, \dodoi{10.3847/1538-4357/834/2/205}

\bibitem[{{Mansheim} {et~al.}(2017{\natexlab{b}}){Mansheim}, {Lemaux},
  {Tomczak}, {Lubin}, {Rumbaugh}, {Wu}, {Gal}, {Shen}, {Dawson}, \&
  {Squires}}]{Mansheim2017hiz}
{Mansheim}, A.~S., {Lemaux}, B.~C., {Tomczak}, A.~R., {et~al.}
  2017{\natexlab{b}}, \mnras, 469, L20, \dodoi{10.1093/mnrasl/slx041}

\bibitem[{{Miller} \& {Owen}(2003)}]{MillerOwen2003}
{Miller}, N.~A., \& {Owen}, F.~N. 2003, \aj, 125, 2427, \dodoi{10.1086/374767}

\bibitem[{{Molnar}(2016)}]{Molnar16review}
{Molnar}, S. 2016, Frontiers in Astronomy and Space Sciences, 2, 7,
  \dodoi{10.3389/fspas.2015.00007}

\bibitem[{{Navarro} {et~al.}(1997){Navarro}, {Frenk}, \& {White}}]{NFW97}
{Navarro}, J.~F., {Frenk}, C.~S., \& {White}, S.~D.~M. 1997, \apj, 490, 493,
  \dodoi{10.1086/304888}

\bibitem[{{Newman} {et~al.}(2013){Newman}, {Cooper}, {Davis}, {Faber}, {Coil},
  {Guhathakurta}, {Koo}, {Phillips}, {Conroy}, {Dutton}, {Finkbeiner}, {Gerke},
  {Rosario}, {Weiner}, {Willmer}, {Yan}, {Harker}, {Kassin}, {Konidaris},
  {Lai}, {Madgwick}, {Noeske}, {Wirth}, {Connolly}, {Kaiser}, {Kirby},
  {Lemaux}, {Lin}, {Lotz}, {Luppino}, {Marinoni}, {Matthews}, {Metevier}, \&
  {Schiavon}}]{Deep2:2013}
{Newman}, J.~A., {Cooper}, M.~C., {Davis}, M., {et~al.} 2013, \apjs, 208, 5,
  \dodoi{10.1088/0067-0049/208/1/5}

\bibitem[{{Sobral} {et~al.}(2015){Sobral}, {Stroe}, {Dawson}, {Wittman}, {Jee},
  {R{\"o}ttgering}, {van Weeren}, \& {Br{\"u}ggen}}]{Sobral2015}
{Sobral}, D., {Stroe}, A., {Dawson}, W.~A., {et~al.} 2015, \mnras, 450, 630,
  \dodoi{10.1093/mnras/stv521}

\bibitem[{{Springel} \& {Farrar}(2007)}]{Springel2007}
{Springel}, V., \& {Farrar}, G.~R. 2007, MNRAS, 380, 911,
  \dodoi{10.1111/j.1365-2966.2007.12159.x}

\bibitem[{{Stroe} {et~al.}(2017{\natexlab{a}}){Stroe}, {Sobral},
  {Paulino-Afonso}, {Alegre}, {Calhau}, {Santos}, \& {van Weeren}}]{Stroe2017}
{Stroe}, A., {Sobral}, D., {Paulino-Afonso}, A., {et~al.} 2017{\natexlab{a}},
  \mnras, 465, 2916, \dodoi{10.1093/mnras/stw2939}

\bibitem[{{Stroe} {et~al.}(2017{\natexlab{b}}){Stroe}, {Sobral},
  {Paulino-Afonso}, {Alegre}, {Calhau}, {Santos}, \& {van
  Weeren}}]{Stroe17Survey}
---. 2017{\natexlab{b}}, \mnras, 465, 2916, \dodoi{10.1093/mnras/stw2939}

\bibitem[{{Stroe} {et~al.}(2015){Stroe}, {Sobral}, {Dawson}, {Jee}, {Hoekstra},
  {Wittman}, {van Weeren}, {Br{\"u}ggen}, \& {R{\"o}ttgering}}]{Stroe15CIZA}
{Stroe}, A., {Sobral}, D., {Dawson}, W., {et~al.} 2015, \mnras, 450, 646,
  \dodoi{10.1093/mnras/stu2519}

\bibitem[{{van Weeren} {et~al.}(2011){van Weeren}, {R{\"o}ttgering}, \&
  {Br{\"u}ggen}}]{RvW2011photometry}
{van Weeren}, R.~J., {R{\"o}ttgering}, H.~J.~A., \& {Br{\"u}ggen}, M. 2011,
  \aap, 527, A114, \dodoi{10.1051/0004-6361/201015991}

\bibitem[{{Wittman}(2019)}]{Wittman19analogs}
{Wittman}, D. 2019, \apj, 881, 121, \dodoi{10.3847/1538-4357/ab3052}

\bibitem[{{Wittman} {et~al.}(2018){Wittman}, {Cornell}, \&
  {Nguyen}}]{Analogs2018}
{Wittman}, D., {Cornell}, B.~H., \& {Nguyen}, J. 2018, \apj, 862, 160,
  \dodoi{10.3847/1538-4357/aacf3e}

\end{thebibliography}
\bibliographystyle{aasjournal}


\section*{Full version of Table 2}

\startlongtable
\begin{deluxetable}{llrc}
  \tablecaption{\Ha\ equivalent width estimates}  
  \tablecolumns{4}
  \tablehead{\colhead{RA} & \colhead{DEC} & \colhead{EW (\AA)} & \colhead{uncertainty (\AA)}}
  \startdata
  00:10:60.0 & 52:32:26.8 & 0.77 & 0.04\\
00:11:05.6 & 52:32:60.0 & 2.21 & 0.17\\
00:11:09.7 & 52:28:42.5 & -26.19 & 0.43\\
00:11:10.0 & 52:28:05.2 & 1.68 & 0.22\\
00:11:10.6 & 52:31:00.1 & 0.88 & 0.11\\
00:11:11.1 & 52:29:50.8 & 1.39 & 0.06\\
00:11:12.2 & 52:31:40.6 & -72.33 & 0.68\\
00:11:13.0 & 52:30:52.1 & -10.34 & 0.69\\
\enddata
\end{deluxetable}

\end{document}